\documentclass[preprint,showpacs,preprintnumbers,amsmath,amssymb,aps,prb]{revtex4}
\usepackage[colorlinks=true,urlcolor=red,linkcolor=blue,citecolor=blue]{hyperref}
 
\usepackage{graphicx}%
\usepackage{dcolumn}
\usepackage{bm}
 
\newcommand{\ia }{\'{\i}}

\begin{document}

\begin{minipage}{15cm}
\centerline{\Large\bf  Graphical Abstract} 

\centerline{\bf Nucleation and Growth of Stellated Gold Clusters: }
\centerline{ \bf Experimental Synthesis and Molecular Dynamics Study}
\centerline{\em J.\ M.\ Cabrera-Trujillo, J.M. Montejano-Carrizales, 
J.L. Rodr\ia guez-L\'opez}
\centerline{\em W. Zhang, J.J. Vel\'azquez-Salazar, and  M. Jos\'e-Yacam\'an}
\vspace*{7cm}
\end{minipage}

\begin{figure} 
\includegraphics[width=\columnwidth]{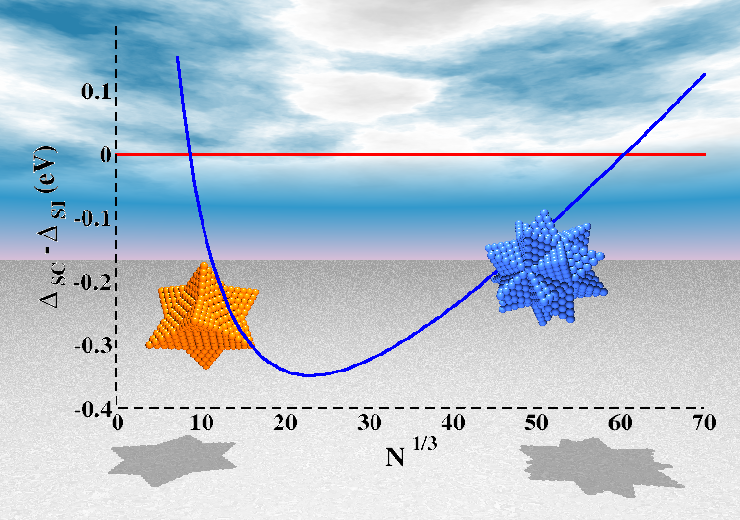}
\end{figure} 

\newpage

\title{Nucleation and 
Growth of Stellated Gold Clusters: Experimental Synthesis and
Molecular Dynamics Study} 

\author{J.\ M.\ \surname{Cabrera-Trujillo}}
\email{cabrera@fc.uaslp.mx}
\affiliation{Facultad de Ciencias, Universidad Aut\'onoma de San Luis
  Potos\'{\i}, 78000 San Luis Potos\'{\i}, Mexico}

\author{J.\ M.\ \surname{Montejano-Carrizales}} 
\affiliation{Instituto de F\'{\i}sica, Universidad Aut\'onoma de San
  Luis Potos\'{\i}, 78000 San Luis Potos\'{\i}, Mexico}  

\author{J.\ L.\ \surname{Rodr\'{\i}guez-L\'opez}}
\email{jlrdz@ipicyt.edu.mx}
\affiliation{Instituto Potosino de Investigaci\'on Cient\ia fica Y
  Tecnol\'ogica, A.C.\ \\ Divisi\'on de Materiales Avanzados; 78216 San Luis Potos\'{\i}, Mexico} 

\author{W.\ \surname{Zhang}, J.J.\ \surname{Vel\'azquez-Salazar}, and  M.\ \surname{Jos\'e-Yacam\'an}} 
\affiliation{Department of Physics and Astronomy, University of Texas
  at San Antonio one UTSA Circle, San Antonio TX 78249 USA}

\date{\today}%

\begin{abstract}
An experimental and  theoretical  study on the structure and energetics
of stellated gold clusters at several sizes is presented.\ 
Systematic molecular dynamics simulations  on Kepler-Poisont 
classified clusters are performed based on cubo-octahedral
 and icosahedral cores, 
following present and  previous 
 studies that suggest stellated clusters grew up 
from these seeds.\ 
  For cluster sizes up to 10802 atoms full atomistic
molecular dynamics simulations at room temperature have been carried
out.\ Results show that stellated clusters in the bigger size regime  
maintain their star-like shape at room temperature, and by means of linear
fitting of energy data to proper function models, 
it is  predicted that the two type of stellated structures
might coexist at room temperature for relatively large sizes, being
these findings in good
agreement with experimental results.\  
\end{abstract}

\pacs{*}

\maketitle
 
%
%

\section{\label{sec:1} Introduction} 
The present study on stellated gold nanoparticles (NPs) is driven 
mainly by fundamental and applied {\em leit motivs}.
From the  fundamental point of view,  
we pursue the hypothesis that stellation on clusters
(Kepler-Poisont solids) is a
process that follows after the nucleation of well defined structures, such
as octahedra and icosahedra (platonic shapes) and cuboctahedra and
truncated octahedra (archimedean solids)\cite{Rodriguez2006}, supported the idea by 
the {\em seed growth approach} in many different syntheses of
anysotropic noble metal
NPs\cite{Burt:2005pd,Senapati2010,Lun2010,Hofmeister2009,Xia2009,Tapan2009}; 
where each one of these
shape classifications  could be 
located in size regions---althought 
with not well defined boundaries---of the  cluster growth.  

From the applied point of view, in the nanoparticle research fields 
there are now well stablished  facts that drive applications researchs in these 
fields, {\em i.e.}, the properties of the systems are influenced by the
NPs size, but also is known now  that the shape strongly 
influences the physical and chemical properties of the
NPs\cite{Murphy2005,Iglesias2006,Noguez2007}. 
 Thus, in the problem of understanding how  shape and size of
metallic  NPs  relate each other, and how these factors influence the
properties of the systems under study,   becomes more complex but 
certainly, more fascinating too. 
Because of the facile synthesis, 
their enhanced absorption and scattering  cross sections that
Au and Ag NPs show under electromagnetic
radiation, and because of their potential nontoxicity, 
noncitotoxicity, and their good biocompatibility, they are widely studied for  
biological imaging and biomedicine applications 
(cancer cell imaging and destruction, photermal therapy, 
among others)\cite{Prashant2006,Huang2006,Wei2009}. 
In particular, Au NPs generate heat under proper electromagnetic 
radiation, being this  heating effect  strong 
 when the energy of the incident 
radiation   is close to the plasmon frequency of the metallic
NPs\cite{Richardson2006}.

Among factors that had contributed to the comprehension, understanding and
development of the  energetics, thermodynamics, and
kinetics of atomic and molecular clusters\cite{Baletto:2005sy}, we can 
mention the interplay between spectroscopy\cite{Burt:2005pd},
and  theoretical techniques based on solid-state 
physics (SSP)\cite{Gupta:1981wb, Tomanek:1985ct, Rosato:1989di, Cleri:1993ck}
or other potentials supported by the density-functional theory 
(DFT)\cite{Finnis:1984vq, Daw:1984jh}. 
The development of geometric models based on platonic
solids\cite{Montejano-Carrizalez:2004wn},  
 or on the most probably structures in experiments generated by  
quenching technique in a molecular dynamics simulation\cite{Stillinger:1982uq} or
by genetic algorithms\cite{Johnston:2003qe} provide us with
suitable geometrical and inter-atomic potential models are available to study small
and large number of particles assemblies through non-quantum
methodologies like Monte Carlo or molecular dynamics simulations (MDS)
at high confident levels.\ 

Under this theoretical framework, extensive studies on metallic
clusters have been carried out\cite{Baletto:2005sy}.\ 
For instance, by using the excess-energy per number of surface atoms
on clusters  as a diagnostic tool, Balletto
{\it et al.}\cite{Baletto:2002ib} had found and explained why   
gold clusters prefer fcc structures at sizes larger than
600 atoms, and strong competitions between D$_h$ and fcc  near 
$N=400$ atoms.\   
Two potential models based on DFT and SSP theories were used in those
calculations, and they also found agreement with other results
obtained at different levels of calculation.\ 
 
Several potential models for transition and noble metals, based on the
tight-binding second-moment approximation (TBSMA)\cite{Gupta:1981wb,
  Tomanek:1985ct}, have been proposed.\ For instance, the
parameterization of Rosato {\it et al.}\cite{Rosato:1989di} (RGL model)
and a second parameterization of the RGL due to Cleri and
Rosato\cite{Cleri:1993ck} (CR), among others\cite{Baletto:2002ib,
  Chamati:2004fj}.\ All these kind of models include the repulsive
pairwise Born-Mayer potential to stabilize the system.\ One of the main
characteristics of those models is the inclusion of many body effects
ensuring correct estimation of several bulk properties, like cohesive
energies, vacancies, surface effects, etc.\ The correct estimation of
those bulk properties, mainly cohesive energy, surface effects, and
surface recostructions, has given confidence to extend the model to
atoms in a cluster environment\cite{Baletto:2005sy}.\ Thus, with these
reliable potential models at hand, general trends can be obtained. 
For example, significative knowledge on the thermal behavior
 of gold clusters is well known for specific
geometries based on crystalline and non-crystalline motifs, like fcc,
icosahedral and decahedral clusters\cite{Baletto:2005sy}.\ 

\begin{figure*}
\includegraphics[width=\columnwidth]{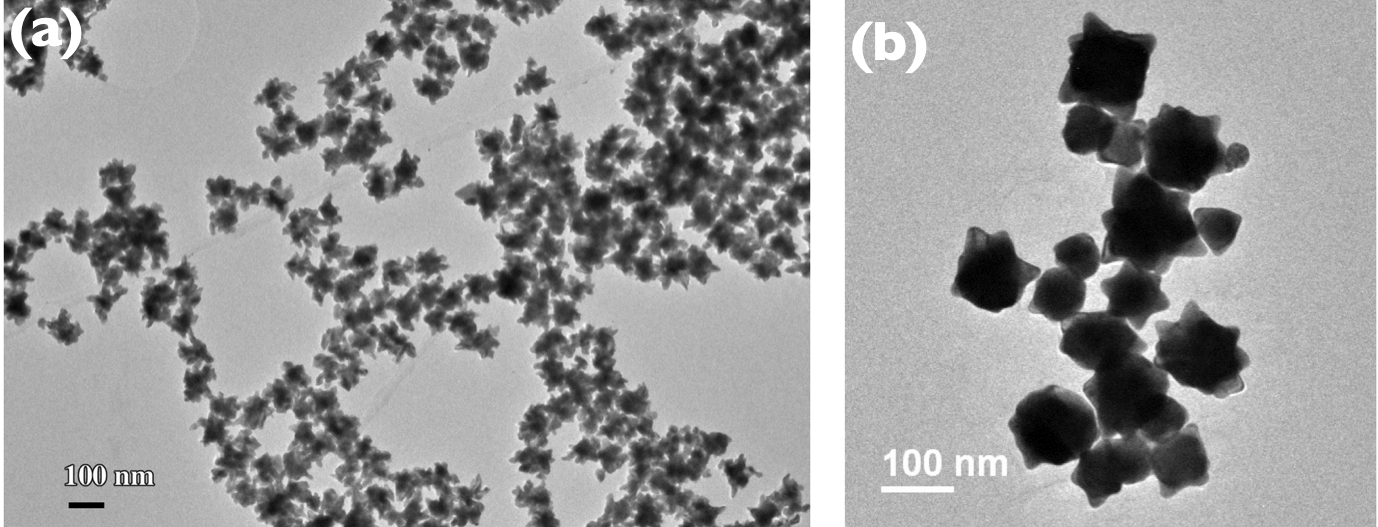}%
\caption{\label{stars}
(a) Low resolution transmission electron microscopy image of
gold nanoparticles with a high production rate of stellated particles.  
In (b) a magnified region where more detail  of these  stellated
clusters is observed.}
\end{figure*}

In this work,  by means of classical MD simulations, and the CR
interatomic potential model, 
we study the structure and dynamics of  stellated gold nanoclusters
with icosahedral and cuboctahedral seeds,  biased mainly  by the 
present experimental results on  gold nanoparticles and previous
reports on stellated (pi\~nata like) clusters\cite{Burt:2005pd}.\ 
Stellated Au nanoparticles have been synthesized at 
room conditions of temperature
and pressure.\ Figure~\ref{stars} shows the great rate of production of these 
stellated clusters, and the high-resolution transmission electron microscopy 
(HRTEM) images at different orientations
clearly show highly symmetric pyramidal structures with slightly
smoothed apices.\ From experimental facts, it has been observed 
high stability in the  clusters shapes, {\em i.e., } they do not change in days and
even months after depositing them in an inert environment.\ These
facts induced the hypothesis that stellated  gold particles 
could have cuboctahedral and  icosahedral cores, but
both with a surface tetrahedral pyramidal growing along the high symmetry
axes.\ In order to give some insight into the structural and thermal
behavior of these stellated structures, mainly at room temperature,
we present a systematic study of the structural and thermal behavior
of two groups of gold clusters whose construction was based on those structures.\

\section{\label{sec:2} Experimental Details}

\subsection{\label{section:2a}Materials and Methods}
Hydrogen tetrachloroaurate trihydrate (HAuCl$_4$ $\cdot$ 3H$_2$O), 
Silver nitrate, trisodium citrate dehydrate, ascorbic acid were purchased from
Sigma-Aldrich and used as received.\ 
Multi-branched polyhedral gold nanoparticles were prepared by colloidal
reduction in aqueous solution at ambient conditions according to previous
literature methods\cite{Burt:2005pd}.\ 
A volume of 0.5 mL of 0.1 M HAuCl$_4$ aqueous solution (50 $\mu$mol)
was rapidly added into an aqueous solution of ascorbic acid (2 mmol in
40 mL H$_2$O) under vigorous stirring.\ The solution immediately
changed its color to opaque orange-red within $1\approx 2$ seconds.\ 
All the samples were purified by washing with ethanol and
centrifuging, and finally dispersed in ethanol.\ 

\subsection{\label{sec:2b} TEM Characterization}
 The Au nanostar samples were characterized by  
ultra-high resolution scanning electron microscope (SEM)
FEG Hitachi S-5500 (0.4 nm at 30 kV)  with BF/DF Duo-STEM detector and
high resolution transmission electron microscope (HRTEM)  Jeol
JEM-2010F FE-TEM with an accelerating voltage of 200 kV, 
 equipped with a Schottky-type field emission gun and  a
 point-to-point resolution of
0.19 nm. Samples for  TEM studies were prepared by
placing directly the gold nanoparticles onto a holey carbon TEM
grid. 

\section{\label{sec:3}Theoretical Background and Methodology}
The functional form of the TBSMA for pure atomic
systems, also known as Gupta potential\cite{Gupta:1981wb},  is given by
\begin{equation}
V(r_{i}) = \sum_{j\neq i}A \exp(-p\,\Delta r_{ij}) - 
\left(\sum_{j\neq i}\xi^2 \exp(-2q\,\Delta r_{ij})\right)^{1/2}\;.
\label{eq:1}
\end{equation}
The first term is the repulsive Born-Mayer pairwise potential, and the
second one is the tight-binding based attractive non-pairwise
potential which takes into account many-body effects.\  
$\Delta r_{ij}=(r_{ij} - r_{0})/r_0$ is the fractional change in
distance between atom $i$ and atom $j$, and $r_{0}$ is the distance
between first nearest-neighbors in bulk.\  
The five parameters, $A, \xi, p, q, \text{ and } r_{0}$ must be
determined for each particular transition or noble metal  atom by a
parametrization procedure using bulk
data\cite{Rosato:1989di,Cleri:1993ck, Lopez:1999yq}.\ If only first
nearest-neighbors are taking into account in Eq.~(\ref{eq:1}), and if
the equilibrium condition is fulfilled, then the number of parameters
is decreasing from five to three\cite{Tomanek:1985ct}: 
\begin{subequations}
 	\label{eq:2}
 	\begin{equation}
 	A=\frac{q}{p-q} \frac{|E_{\textrm{coh}}|}{Z}\label{eq:2.1},
	\end{equation}	
	\begin{equation}
	\xi=\frac{p}{p-q} \frac{|E_{\textrm{coh}}|}{\sqrt{Z}}\label{eq:2.2},
	\end{equation}
\end{subequations}
where $E_{\textrm{coh}}$ is the experimental cohesive energy of bulk
used in the parameterization, and $Z$ is the number of first
nearest-neighbors in the crystal.\ 

\begin{figure*}
 \includegraphics[width=\columnwidth]{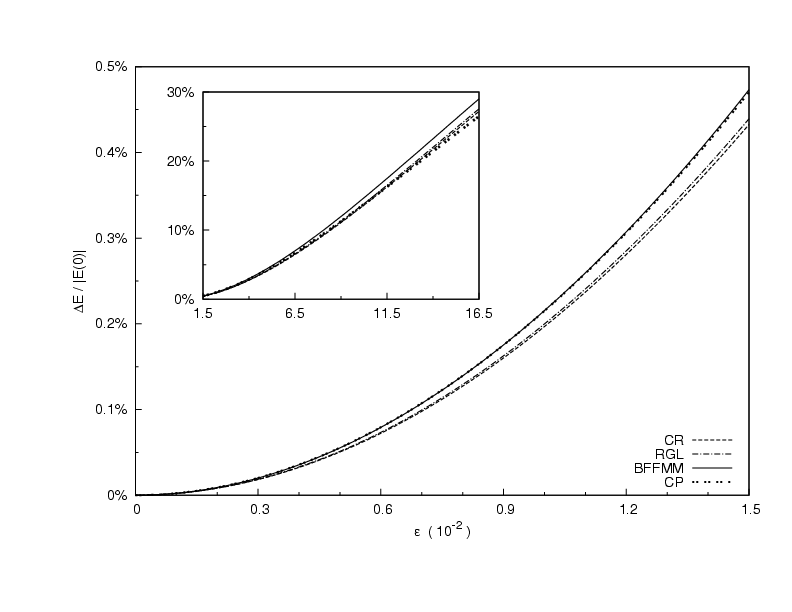} 
\caption{\label{fig:2} 
Behavior of the fractional change in the cohesive energy accompanying
a fractional change in the first nearest-neighbors distance $\epsilon
= (r_{ij} - r_{0})/r_0$ as given by Eq.~(\ref{eq:3b}) for the
potential models given in Table~\ref{tab:1}.\ 
} 
\end{figure*}

Some sets of parameters for gold (fcc crystal) are given in
Table~\ref{tab:1}.\ In these models values for $A$ and $\xi$ are very
close to those calculated from Eqs.~(\ref{eq:2}) using the
experimental value $E_{\textrm{coh}} = -3.78 \text{ eV/atom}$ and $Z =
12$, both from gold fcc--crystal\cite{Kittel:2005uf}.\ 
\begin{table}
\caption{\label{tab:1} Comparision of published potential parameters for gold
  derived by fitting bulk  experimental data to values calculated
  using the Gupta functional form given by Eq.~(\ref{eq:1}).\ We note
  that values for the lattice constant and cohesive energy,
  $(\sqrt{2}\ r_{0},\ E_{\textrm{coh}})$, in each model are
  experimental values taken from Kittel\cite{Kittel:2005uf}, but on
  different potential versions: the
parameterization of Rosato {\it et al.}\cite{Rosato:1989di} (RGL model,
Kittel 1972); the second parameterization of the RGL due to Cleri and
Rosato\cite{Cleri:1993ck} (CR, Kittel 1966); whereas  Balleto {\em et
  al.}\cite{Baletto:2002ib} (BFFMM) and Chamati and
Papanicolaou\cite{Chamati:2004fj} (CP) use parameters 
from Kittel, 1996.\  
For the CP model we have used 
the calculated values\cite{Chamati:2004fj} shown in the row.} 
\begin{ruledtabular}
\begin{tabular}{ccccccc}
 Model\footnotemark[1]& $p$	& $q$ & $\xi$ (eV) & $A$ (eV) & 
$\sqrt{2}~r_{0}$ (\AA) & $-E_{\textrm{coh}}$ (eV/atom) \\
\hline 
 RGL\cite{Rosato:1989di}
 & 10.15	& 4.13 & 1.8398\footnotemark[2]	
& 0.2161\footnotemark[2] & $4.08$ & 3.78 \\
 
 CR\cite{Cleri:1993ck}
 & 10.229 & 4.036 & 1.790 & 0.2061 & $4.079$ & 3.779 \\
 
 BFFMM\cite{Baletto:2002ib}
 & 10.53	& 4.30 & 1.855 & 0.2197 & $4.07$ & 3.78 \\
 
 CP\cite{Chamati:2004fj}
 & 14.6027 & 3.12572	& 1.32795 & 0.08170 & $4.068$ & 3.76 
 
\end{tabular}
\end{ruledtabular}
\footnotetext[1]{Model's name is proposed from the initials of the
  author and coauthors of the cited article.} 
\footnotetext[2]{Values obtained using Eqs.~(\ref{eq:2}).}
\end{table}
At a first sight, differences and similarities between these potential
models can be obtained from the fractional change in total energy
accompanying a fractional change in distance between first
nearest-neighbors $\epsilon = (r_{ij} - r_{0})/r_0$, which at second
order approximation in $\epsilon$ is roughly given by 
\begin{equation}
\frac{ \Delta E}{|E(0)|} \approx \sigma_{1}\ \epsilon^{2}
\label{eq:3a},
\end{equation}
where $\Delta E = E(\epsilon)-E(0)$, $E$ is the total energy, and
$\sigma_1=pq/2$, see Ref.~\onlinecite{Baletto:2002ib}.\  
The $\sigma_1$ values for the CR and RGL models differ $1.5\%$ each
other, and the values among BFFMM and CP differ $0.8\%$; 
corresponding the lower and bigger values to the CR and
CP models, respectively.\ Thus, according to $\sigma_{1}$ values and
Eq.~(\ref{eq:3a}),  the BFFMM potential is as sticky as the CP model,
and the CR model is slightly less sticky than the RGL, being CP more
sticky than the CR model.\ However, according to the formula 
\begin{equation}
\frac{ \Delta E }{ | E(0) | } = 
\frac{ 1 }{ p-q }( q ( e^{-p\epsilon}-1 ) - p( e^{-q\epsilon}-1 ) )
\label{eq:3b},
\end{equation}
where no approximations in $\epsilon$ have been used, the similarity
between the BFFMM and CP models only occurs for displacements less
than $1.5\%$ of $r_{0}$, as can be seen from the plots in
Fig.~\ref{fig:2}.\ In fact, crossovers between the CP and the BFFMM,
RGL and CR models occur at $\epsilon \simeq 0.0085$, $\epsilon \simeq
0.093$, and $\epsilon \simeq 0.115$ respectively.\ Thus, according to
Eq.~(\ref{eq:3b}) the CP model is less sticky than the other
potentials, but for $\epsilon \gtrsim 0.115$.\          

\subsection{\label{sec:3a}Simulation Details}
In order to perform MD simulations at room temperature of the SC and
SI gold clusters described above, the  Au potential
model given by Eq.~(\ref{eq:1}) was used, applying  the four parameter
sets given in 
Table~\ref{tab:1} under the GROMOS96 scheme\cite{Scott:1999kx}.  
This package applies  the leap-frog algorithm to integrate the 
Newton's classical equations of motion under vacuum or
periodic boundary conditions\cite{Allen:2002jy}.\ High
numerical stability is ensured for a time-step of 2 fs, if shake
method is applied; otherwise 0.5 fs.\ Those time-steps values are
commonly used for flexible biological molecules\cite{Scott:1999kx};
for metallic atoms a practical value for the time-step is 1
fs\cite{Wilson:2000bv}.\ Also under this scheme, the system could be
weakly or strongly coupled to a temperature bath using the Berendsen's
thermostat\cite{Berendsen:1984to}.\  
The numerical ensemble produced by MDS applications using Berendsen's
thermostat is the weak-ensemble\cite{Morishita:2000qv}, which has
intermediate property between the canonical ($N,V,T; \alpha=0$) and
microcanonical ($N,V,E; \alpha=1$) ensembles.\  
Here, $\alpha$ is a constant in equilibrium but dependent on the
temperature coupling
parameter\cite{Berendsen:1984to,Morishita:2000qv}.\ 
In the practice,  $\alpha$ is determined from the ratio of standard deviations of
kinetic and potential energies, 
\begin{equation}
\alpha = \sqrt{ \left< \left( \delta E_{\textrm{Kin}} \right) ^{2} 
\right> / \left< \left( \delta E_{\textrm{Pot}} \right) ^{2} \right>}
\label{eq:5},
\end{equation}
where $\delta E_{z}(t) = E_{z}(t) - \left<E_{z}\right> $ stands for
the instantaneous fluctuation of property $z$ (kinetic or potential
energy), and $\left<\;\right>$ denotes average time.

\subsection{\label{sec:3b} Geometrical Description}
Since any stellated gold clusters
considered in this work are made up of one icosahedral or one
cuboctahedral cluster in the core,  and regular tetrahedral clusters attached to
their (111) faces,  we will refer them as stellated icosahedral (SI)
and stellated cuboctahedral (SC) clusters, respectively.\     
Specifically, each SI gold cluster is made up of one Mackay
icosahedral cluster with ($n=k+1$, $k\geqslant 1$) shells and $N(n)
= \frac{10}{3}n^{3}-5n^{2}+\frac{11}{3}n-1$ atoms, and twenty
tetrahedral clusters, each with $N_{\textrm{Td}}(k)=
\frac{k(k+1)(k+2)}{6}$ atoms and attached to each one of the twenty
(111) facets of the icosahedral cluster.\   
In a similar way, each SC gold cluster is made up of one cuboctahedral
cluster with $N(k+1)$ atoms, and eight tetrahedral clusters, each with
$N_{\textrm{Td}}(k)$ atoms and attached to each one of the eight (111)
facets of the cuboctahedral cluster.\  
Hence, the number of atoms of any ${\textrm{SI}_{k}}$ cluster is given
by $N_{\textrm{SI}}(k) = N(k+1) + 20N_{\textrm{Td}}(k)$, while
$N_{\textrm{SC}}(k) = N(k+1) + 8N_{\textrm{Td}}(k)$ is the number of
atoms of any ${\textrm{SC}_{k}}$ cluster.\ $N_{\textrm{SI}}(k)$ and
$N_{\textrm{SC}}(k)$ are only different in the number of tetrahedrons,
and they can also be expressed as  
\begin{subequations}
 	\label{eq:4}
 	\begin{equation}
 	N_{\textrm{SC}}(k) = \frac{14}{3}k^{3} +  9k^{2} + \frac{19}{3}k+1\label{eq:4.1},
	\end{equation}	
	\begin{equation}
	N_{\textrm{SI}}(k) = \frac{20}{3}k^{3} + 15k^{2} + \frac{31}{3}k+1\label{eq:4.2},
	\end{equation}
\end{subequations}
with $k\geqslant 1$.

The gold clusters considered in this work are those SC clusters with
$k \leqslant 13$, and those SI with $k \leqslant 11$, shown in
Fig.~\ref{fig:3}.\ In all the initial geometrical models the distance
between first nearest-neighbor atoms is fixed to the first
neares-neighbor distance of bulk gold, $r_0~=~2.88 \text{ \AA}$.\

\section{\label{sec:4}Results and Discussion} %
Figure~\ref{fig:3} displays the unrelaxed stellated clusters used in 
this study.\ To prepare these SI and SC gold clusters 
for MD simulations at $T = 298 \text{ K}$, we have first slightly
relaxed each one of them with the steepest-descents method using  the four
potential models in Table~\ref{tab:1}.  
Although the main purpose of the minimization procedure is to reduce
the internal strain of the cluster, and the consequently reduction of
computer effort at the equilibration period, we have analyzed these
results to gain insight investigating how the residual stress is
reflected into 
the structures,  
 how the surface and volume contributions to the total energy behaves
 as function of cluster size, and what type of the stellated
structures is the most favorable. 
Similarities and differences between the results obtained using older
and newer gold potential models (Table~\ref{tab:1}), are also
enhanced and analyzed at this stage in order to choose the appropriate
potential to perform the simulations at room temperature. 
Thus, a conciencious analysis of the structural and energetic results obtained
under this procedure is presented in Sec.~\ref{sec:4a}.\ In
Sec.~\ref{sec:4b}, we have posed similar questions for the
equilibrated clusters at $T = 298 \text{ K}$, and the results have
been analyzed in the same way, but only for the CR potential model,
which was elected after the analysis 
in Sec.~\ref{sec:4a}.     

\subsection{\label{sec:4a}Relaxed Clusters} %

In cluster physics is common to describe atoms as hard spheres, and to
use them as a linear lengt unit. Following this idea,   in  the micro-crystal size regime  
(where a high rate  of the total  $N$ atoms belong to  the bulk)
the particles are   $N^{1/3}$ atoms long, with area facets
proportional to $N^{2/3}$ atoms. Only some
atoms belong to the surface ($\sim N^{2/3}$),  
located predominately on the facets and only  
a negligible number  on the edges and the vertices. On the contrary, when the
cluster size is decreased, almost all the atoms tend to be 
at the surface. In this case, however, the amounts of atoms in the edges, the
vertices, and the facets tend to compete among them, and the extreme
case is for the very smaller clusters that contains only vertex
atoms. Therefore, it would be helpful to obtain a function that give
insights  in which range of $N$, the macro/micro ($N_{\textrm{surf}} \ll N $)
or the nano ($N_{\textrm{surf}} \approx N $) character of any studied
cluster is manifested. 

For this purpose, it is  proposed a function for the number
of surface atoms $N_{\textrm{surf}}$,  in terms of $N^{1/3}$, $N^{2/3}$,
 and a constant, as follows   
 
\begin{equation} 
  N_{\textrm{surf}}(N) = \alpha N^{2/3}+ \beta N^{1/3} + \gamma
  \label{eq:6},
\end{equation}
where the coefficients $\alpha,~\beta, \text{ and } \gamma$
must be free parameter, since the function must describe the full
number of surface atoms for small, intermediate, and large size
clusters. Thus, when fitting the function to data a high 
correlation between them is expected and some of them could get negative values. 

The behavior of the computed ratio, $N_{\textrm{surf}}/N^{2/3}$, as a 
function of the cluster size, is almost the same for both the relaxed and
unrelaxed SI and SC clusters, as is shown in Fig.~\ref{fig:4}.   
The number of surface atoms of each relaxed cluster,
$N_{\textrm{surf}}$, has been calculated through the Lee--Richard
algorithm\cite{Lee:1971bx} using the Van der Waals radius of gold as a
probe radius $r_{\textrm{probe}} = 1.66$~\AA. 
Hereafter we denote this function as ${\cal N}(N)=N_{\textrm{surf}}/N^{2/3}$ 
in order to facilitate discussions. 
In the Figure, the solid and dotted lines are obtained by linear
fitting of SC and SI $N_{\textrm{surf}}$ data to functions of the form
given by Eq.~(\ref{eq:6}), respectively. 
Here the function ${\cal N}_{\textrm{SC}}(N)$ is for $N \geqslant 21$,
while ${\cal N}_{\textrm{SI}}(N)$ is for $N \geqslant 135$.  
\begin{table}
\caption{\label{tab:2}The parameters for the fitting function,
  Eq.~(\ref{eq:6}), together with its asymptotic standard errors for
  the SC and SI gold clusters relaxed with the steepest-descents
  and the CR potential model.} 
\begin{ruledtabular}
\begin{tabular}{cccc}
    & $\alpha$ & $\beta$ & $\gamma$ \\
\hline
 SC & $6.4490  \pm 0.0005  $ & $-13.97  \pm 0.01    $ & $9.33    \pm 0.08    $ \\
 SI & $8.4663  \pm 0.0006  $ & $-39.74  \pm 0.02    $ & $73.15   \pm 0.11    $ 
\end{tabular}
\end{ruledtabular} 
\end{table}
The coefficients of the fitting functions for the SC and SI clusters
are given in Table~\ref{tab:2}.  
Note that some parameters are necessarily negative.

As can be seen from Fig.~\ref{fig:4}, 
calculated values for SC clusters tend faster to its asymptotic value,
$\alpha_{\textrm{SC}} = +6.449$, than those values for SI clusters,
that tend to $\alpha_{\textrm{SI}} = +8.466$. 
The asymptotic value $\alpha_{\textrm{SC}}$ is greater than the
asymptotic value for a simple cubic cluster, $\alpha_{\textrm{cubic}}
= 6.0$, since the $SC$ clusters are based on a truncated fcc with
tetrahedral units attached to the (111) faces. The value
$\alpha_{\textrm{SI}} = +8.466$ is explained by a similar reasoning.  
For cluster sizes of $\textrm{O}(10^{6})$,  
values of the fitting functions ${\cal N}_{\textrm{SC}}$ and ${\cal
  N}_{\textrm{SI}}$ are around the $2.2\%$ and $4.6\%$ of their
asymptotic values, respectively.\  
Thus, one could expect stronger surface effects for SI clusters than
for the SC clusters, or one could expect that the SC clusters already
behave as macro clusters ($N_{\textrm{surf}}\propto N^{2/3}$) for
relatively small sizes, $N\sim10^{6}$, so that many properties of SC
could be dominated by volume atoms.\ For SI clusters in the same range one
could expect properties dominated by surface atoms.\ As a note of
caution, the stellated clusters have very specific sizes, see
Eqs.~(\ref{eq:4}), because they are constructed  using complete
tetrahedral units mounted on the eight triangular faces of
the cuboctahedra or over the twenty faces  of the icosahedral clusters.\ Hence,
detailed interpolation or extrapolations results from Eq.~(\ref{eq:6})
must be taken with care.\ 

In the small size range, the stellated clusters with sizes
$N_{\textrm{SC}}(4)=469$ and $N_{\textrm{SI}}(4)=709$ contain
tetrahedral units of 20 atoms, and  interesting is to note that
tetrahedral gold cluster with 20 atoms have been proven to be the more
stable gold isomer\cite{Fernandez:2004rm}.\ 
To this respect, we also note that after the slightly relaxation
processes, where each of the potential models displayed in
Table~\ref{tab:1} were used, we have observed that the shape of each
cluster in Fig.~\ref{fig:3} is maintained; the relaxed clusters are
contracted in the average, as it is expected for these kind of
potential models\cite{Gupta:1981wb}, being more evident 
this result in the smaller ones due to
the high value of the ${\textrm{surface}}/{\textrm{volume}}$ ratio,
but the peaks of all the clusters are maintained, which can be
attributed to the same mechanism that maintains vertex atoms of the
global minima tetrahedral cluster of Fernandez {\it et
  al.}\cite{Fernandez:2004rm}.\

In order to explore the energetic trends of the relaxed SC and SI clusters
under the potential models given in Table~\ref{tab:1}, we have first fitted
the potential energy data of each series to a function of the
form (see Ref.\ \onlinecite{Baletto:2005sy} and references therein)  
\begin{equation}
E(N) = A N + B N^{2/3} + C N^{1/3} + D
\label{eq:9},
\end{equation}
where the coefficient $A$ is proportional to the energy per atom of
the bulk-like atoms, that follows from the asymptotic behavior of
$E(N)/N$ for large clusters. The number of bulk atoms  
is taken proportional to the cluster size $N$. 
Thus, at least for large clusters, the first term in the Equation could
be interpreted as the contribution to total energy due to bulk
atoms, and the last three term together as the contribution to total
energy due to the atoms at the surface. 
The explicit dependence on $N^{n}~(1/3,~2/3,\text{ and } 0)$ comes from
the splitting of the full number of surface atoms into facets atoms
(proportional to $N^{2/3}$), edges atoms  (proportional to $N^{1/3}$),
and vertexes atoms, a constant.
 
Therefore, by construction, the first term in Eq.~(\ref{eq:9}) is
interpreted as the total volume contribution to the energy, and the
second, third, and fourth terms together are interpreted as the total
surface energy contribution (see  Ref.\ \onlinecite{Baletto:2005sy} for details).  
As in the case of $N_{\textrm{surf}}(N)$, see Eq.~(\ref{eq:6}), the
coefficients in Eq.~(\ref{eq:9}) must be free parameters, and high
correlation among them are expected (mainly among coefficients B, C,
and D) when fitting data of small, intermediate or large size clusters
to this equation.  

\begin{table}
\caption{\label{tab:3}The coefficients for the fitting function,
  Eq.~(\ref{eq:9}), for the steepest-descents relaxed SC (first row) and
  SI (second row) clusters with the potential models given in
  Table~\ref{tab:1}.\ Units are eV/atom.} 
\begin{ruledtabular}
\begin{tabular}{cccccc}
	Model & $A$& $\sigma$\footnotemark[1]& $B$ & $C$ & $D$ \\
\hline

  CR  & $-3.7795 \pm 0.0003  $ & $0.0000  $ & $1.570   
\pm 0.007   $ & $-0.71   \pm 0.06    $ & $1.68    \pm 0.11    $ \\
      & $-3.7687 \pm 0.0002  $ & $0.0108  $ & $1.678   
\pm 0.003   $ & --    & $-3.53   \pm 0.05    $ \\

  RGL & $-3.8256 \pm 0.0002  $ & $0.0000  $ & $1.454   
\pm 0.007   $ & $-0.73   \pm 0.05    $ & $1.76    \pm 0.10    $ \\
      & $-3.8160 \pm 0.0002  $ & $0.0096  $ & $1.554   
\pm 0.003   $ & --    & $-3.21   \pm 0.04    $ \\

BFFMM & $-3.8284 \pm 0.0002  $ & $0.0000  $ & $1.436  
 \pm 0.005   $ & $-0.77   \pm 0.04    $ & $1.84    \pm 0.08    $ \\
      & $-3.8186 \pm 0.0002  $ & $0.0099  $ & $1.533   
\pm 0.004   $ & --    & $-3.16   \pm 0.05    $ \\

   CP & $-3.7600 \pm 0.0001  $ & $0.0000  $ & $2.967   
\pm 0.004   $ & $-0.54   \pm 0.03    $ & $0.34    \pm 0.06    $ \\
      & $-3.7336 \pm 0.0013  $ & $0.0264  $ & $3.125   
\pm 0.021   $ & --    & $-7.87   \pm 0.29    $ 

\end{tabular}
\end{ruledtabular}
\footnotetext[1]{$\sigma = A - E_{\textrm{ref}}$, 
where $E_{\textrm{ref}} = A_{\textrm{SC}}$.} 
\end{table}

The parameters for the fit to each type of stellated clusters and
potential models are given in Table~\ref{tab:3}.\ 
To avoid large asymptotic standard errors in the linear fitting data
of SI clusters, the coefficient C is excluded from its fitting
process.\ Note that, for the potential models RGL and BFFMM, the
asymptotic values for the binding energy, $A_{\textrm{SC}}$, are lower
than the experimental cohesive energy of bulk used in the
parameterization of each of the potential model (see
Table~\ref{tab:1}), however they are very close to the experimental
cohesive energy, $E_{\textrm{coh}}=-3.81 \text{ eV/atom}$ reported in
Ref.~\onlinecite{Kittel:2005uf1}.\  
For the CP model $A_{\textrm{SC}}$ is exactly the calculated value
of the cohesive energy of bulk Au reported by Chamati and
Papanicolaou\cite{Chamati:2004fj}, and for the CR model the value of
$A_{\textrm{SC}}$ is very close to the experimental cohesive energy of
bulk used in its  parameterization\cite{Cleri:1993ck}.\  

\begin{figure*} 
\includegraphics[width=\columnwidth]{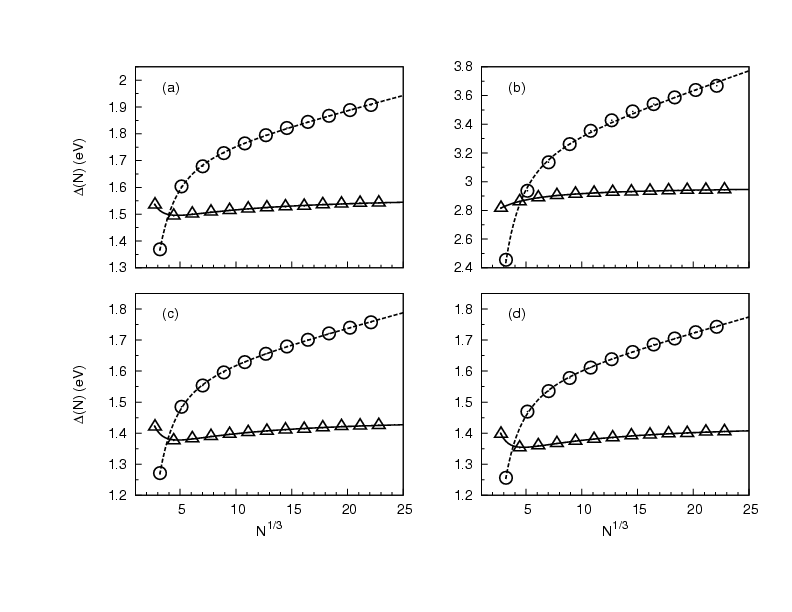}
\caption{\label{fig:5} 
Trends of the calculated excess energy per $N^{2/3}$, $\Delta(N)$, as
function of $N^{1/3}$, for both SC ($\triangle$), and SI ($\bigcirc$)
relaxed clusters for the potential models CR (a), CP (b), RGL (c), and
BFFMM (d).\ The solid and dotted lines are the curves
$\Delta_{\textrm{SC}}(N)$, and $\Delta_{\textrm{SI}}(N)$ from
Eq.~(\ref{eq:8}), where the coefficients are given in
Table~\ref{tab:3}, respectively.\ For each potential model
$E_{\textrm{ref}}=A_{\textrm{SC}}$ has been used.} 
\end{figure*}

In order to explore which type of relaxed structure is more favorable,
we have used as diagnostic tool the excess energy per $N^{2/3}$,
defined by\cite{Baletto:2005sy}, 
\begin{equation}
{\Delta}(N) = \frac{E - N E_{\textrm{ref}}}{N^{2/3}}
\label{eq:7};
\end{equation}
where $E$ stands for the total potential energy of one cluster of size
$N$, and it is common to take for the reference energy
$E_{\textrm{ref}}$ the experimental cohesive energy of the bulk used
in the parameterization of the potential model, see
Table~\ref{tab:1}.\ 
Here, for each potential model, $\Delta(N)$ is calculated using 
$E_{\textrm{ref}}=A_{SC}$, where $A_{SC}$ is given in
Table~\ref{tab:3}.\ 
The results of the calculations of $\Delta(N)$ using each of the
potential models in Table~\ref{tab:1} and the steepest-descents method
are given in Fig.~\ref{fig:5}.\  
It can be observed from this figure that st this stage of the
simulations, almost all SC clusters, except $N_{\textrm{SC}}(1)=21$,
are more favorable than the SI clusters.\  
This behavior occurs for all potential models, except the 
CP model which also excludes $N_{\textrm{SC}}(2)$.\ 
Similar results have been found by Baletto et
al.\cite{Baletto:2002ib}, using the RGL and embedded-atom models, but
for non-stellated icosahedral and cuboctahedral clusters.\  
However, as can be seen from the calculated data in the figure, the
effects of mounting tetrahedral motifs on the triangular faces of
cuboctahedral or icosahedral clusters changes significantly the
behavior of $\Delta(N)$.\  
For instance, $\Delta$ data of icosahedral clusters in
Ref.~\onlinecite{Baletto:2002ib} is concave up, while the $\Delta$
data of   SI clusters in Fig.~\ref{fig:5} is concave down.\ For
Lennard-Jones icosahedral and fcc cluster
data\cite{Xie:612,Northby:1989yq},  it has been found that $\Delta$ is
concave down for both types of clusters.\ We have explored the
concavity effect by expressing $\Delta(N)$ in the form  
\begin{equation}
{\Delta}(N) = \sigma N^{1/3} + B + C/N^{1/3} + D/N^{2/3}
\label{eq:8}~,
\end{equation}
which follows using Eq.~(\ref{eq:9}) into Eq.~(\ref{eq:7}).\ The
parameter $\sigma = A - E_{\textrm{ref}}$ is interpreted as the strain
or elastic energy per atom of the 
cluster\cite{Northby:1989yq,Soler:2000uq,Baletto:2005sy}.\  
It is well known that for clusters with fcc symmetry, 
$\sigma_{\textrm{fcc}}$ is negligible since the bulk energy per
atom $A_{\textrm{fcc}}$,  is the optimum.\ On the contrary, it is also
well known that the non-crystalographic symmetry of the clusters
frustrates the elimination of strains by the minimization process, and
consequently they can accumulate that energy as
elastic-energy\cite{Northby:1989yq,Soler:2000uq,Baletto:2005sy},
which increases with cluster size.\ 
In principle, $A$ must be higher or equal to the cohesive energy of
the bulk crystal, $E_{\textrm{coh}}$, which can be appreciated as the
optimum energy per atom of volume atoms; however in practice, when
fitting MDS results to Eq.~(\ref{eq:9}), $A$ could be slightly lower
than $E_{\textrm{coh}}$.\  
This could be attributed to the linear fitting of MDS data, and or to
the cut-off radius used in the interactions: in all potential models
we have included more than one coordination shell in the interactions.\  
In any case, since one is interested in the relative stability of the
two types of clusters,  
to avoid negative values for $\sigma$ in Eq.~(\ref{eq:8}) we have used
for the reference energy $E_{\textrm{ref}}$ the lower value for $A$ in
each potential model, instead of the experimental cohesive energy of
bulk used in the parameterization of the potential model, see
Table.~\ref{tab:1}.\ 

In Fig.~\ref{fig:5}, the solid and dotted lines in each plot are the
curves $\Delta_{\textrm{SC}}(N) \text{ and } \Delta_{\textrm{SI}}(N)$
obtained from Eq.~(\ref{eq:8}) with the coefficients given in
Table~\ref{tab:3}, for the potential models CR (a), CP (b), RGL (c),
and BFFMM (d).\  
Note that, in general, the asymptotic value for the binding energy
$A_{\textrm{SC}}$, is lower than the experimental cohesive energy of
bulk used in the parameterization of the potential model, see
Table~\ref{tab:1}.\  
As pointed out in the preceding paragraph, to avoid negative values of
$\sigma_{\textrm{SC}}$, we have used $E_{\textrm{ref}} =
A_{\textrm{SC}}$ in the Eq.~(\ref{eq:8}) for the SC and SI
clusters.\ This procedure has not effect in the present analysis, since only we
have changed the reference point of energy.\ For all the potential models the
respective functions, $\Delta_{\textrm{SC}}(N)$ and
$\Delta_{\textrm{SI}}(N)$, predict only one crossover size.\ 
Only the crossover sizes for the RGL, BFFMM, and CR models agree with
the rule: the larger is $pq$, the smaller are the crossover
sizes\cite{Baletto:2002ib, Baletto:2005sy}.\ The CP model has the
higher value of $pq$, and also the 
higher value of the crossover size.\ 
The exact values of the crossover sizes have no real significance,
 since the calculations have been limited to clusters sizes
given in Eqs.~\ref{eq:4}.\   
We only could infer from these results that both types of clusters
could coexists around these sizes.\  
For any potential model, no others crossover sizes occur for the
larger clusters of both types, since their asymptotes, right lines in
$\Delta(N^{1/3})$, are not intersected.\ 
Thus, each one of the potential models predict only one crossover size,
which occurs for relatively small cluster sizes.\ Therefore, at this stage
of the MD simulations we could partially conclude that (a) all the
slightly relaxed structures maintain the star-like shape of the
initial structures; (b) all SC clusters, except the smaller cluster,
are more favorable than the SI clusters; (c) the most strained SI
structures are given by the CP model, and the less strained are given by
the RGL model (elastic energy values given by the CR and BFFMM models
are very close to the value given by the RGL model); and (d) only the
asymptotic value for the binding energy $A_{\textrm{SC}}$, given by
the CR model, is very close to the value of the experimental cohesive
energy used in its parameterization.\ 
  
\subsection{\label{sec:4b}Equilibrated Clusters at $T=298$ K} 
MD simulations at room temperature of the relaxed clusters  
described above have been conducted as follows: for each cluster a
sample of mechanical states of length 0.25 ns has been produced in a
weak-ensemble approximating the microcanonical-ensemble ($\alpha
\simeq 1$),  and
after $2$ ns of equilibration length.\  
This procedure has been applied to each cluster in the range $21
\leqslant N \leqslant 3095$, including $N=5 631 \text{ and } N=6169$,
and except those with sizes $N_{\textrm{SC}} = 9437$ and
$N_{\textrm{SI}} = 10803$, where an equilibration length of $1$ ns was
used.\ 
In the equilibration period, the Berendsen's thermostat with
time-coupling parameter $\tau_{T} = 0.4$ ps and target temperature   
$T_{0}=298$ K was used\cite{Berendsen:1984to,Morishita:2000qv}.\ 
In all the simulations a time-step of 1 fs is used, and periodic
boundary conditions has been applied, but with a much larger box
length than the cluster radius to ensure non-interacting images.\  
A cutoff radius equal to half box length was used for clusters sizes
$N \leqslant 87$, otherwise a cutoff radius of 2 nm was set.\  
Statistical properties of the produced sampled distribution 
were obtained using the coarse-graining stratified systematic sampling
dividing the sample in equal bins of length $0.05$ ps, to avoid serial
correlations.\ The averages of the properties considered in this work
are reported at the $95\%$ of confident level.\ All  simulation
results have been obtained using only the CR potential model, since
the asymptotic behavior of $E(N)/N$ for the relaxed SC clusters
reproduces very well the experimental cohesive energy used in the
parameterization of the model.\ 

\begin{figure*}
\includegraphics[width=\columnwidth]{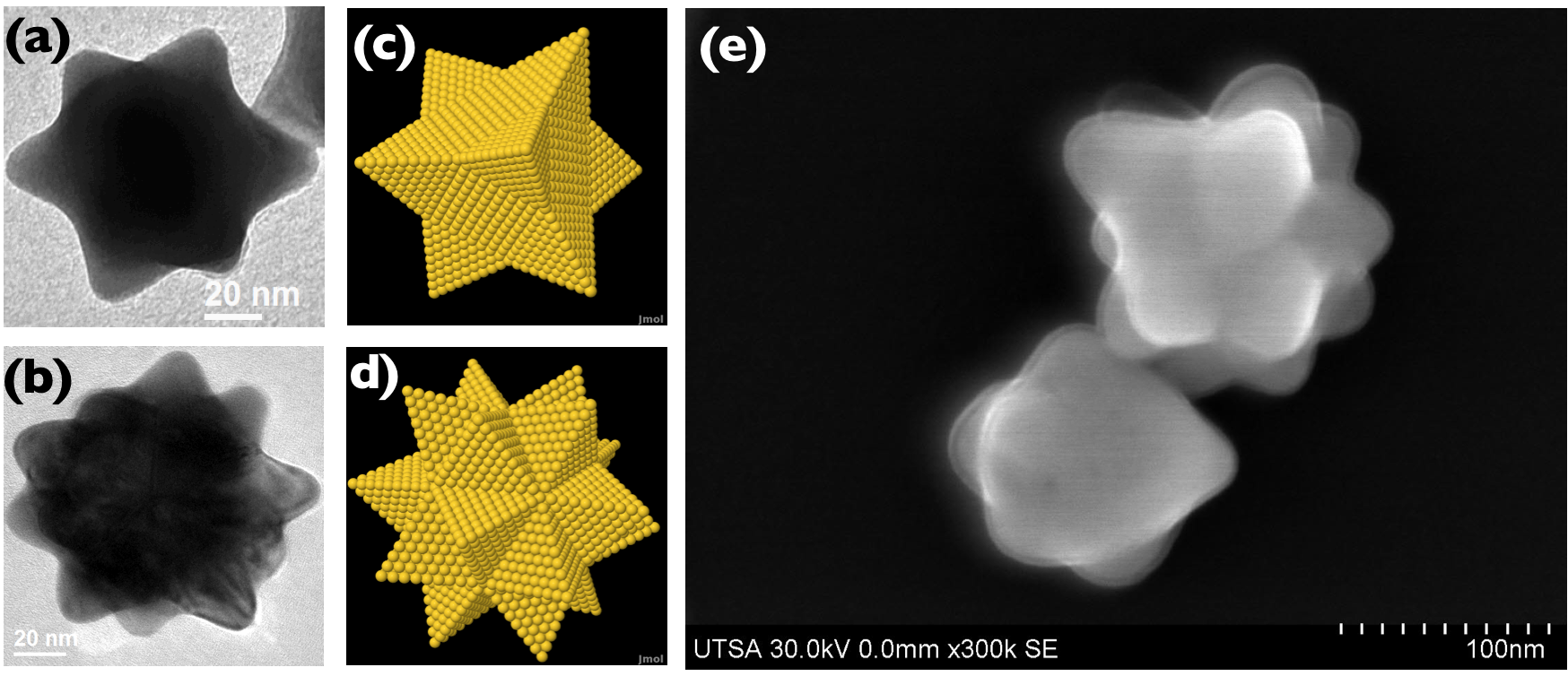}%
\caption{\label{fig:1}
Transmission electron microscopy images of gold particles with assumed
 (a) cuboctahedral and (b) icosahedral seeds.\ Models that resemble
 these images are shown in (c) and (d).\ SEM image in (d) evidence 
both kind of  stellated clusters coexisting.\   
}
\end{figure*}

\begin{figure*} 
\includegraphics[width=\columnwidth]{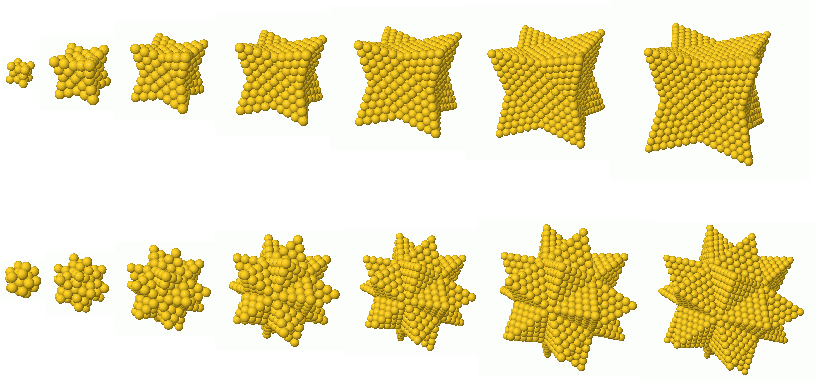} 
\caption{\label{fig:3}
Some star-like gold clusters, similar to those shown in
Fig.~\ref{fig:1}, where each one was obtained by using regular
tetrahedral motifs attached to the eight (111) faces of a
cuboctahedron (top row) or the twenty (111) faces of the icosahedron
(bottom row), both with magic size.\ 
These pi\~nata-like clusters correspond, from left to right, to $k=1, 2, 3,...,7$, see
Eqs.~(\ref{eq:4}).\ 
} 
\end{figure*}

\begin{figure*} 
\includegraphics[width=\columnwidth]{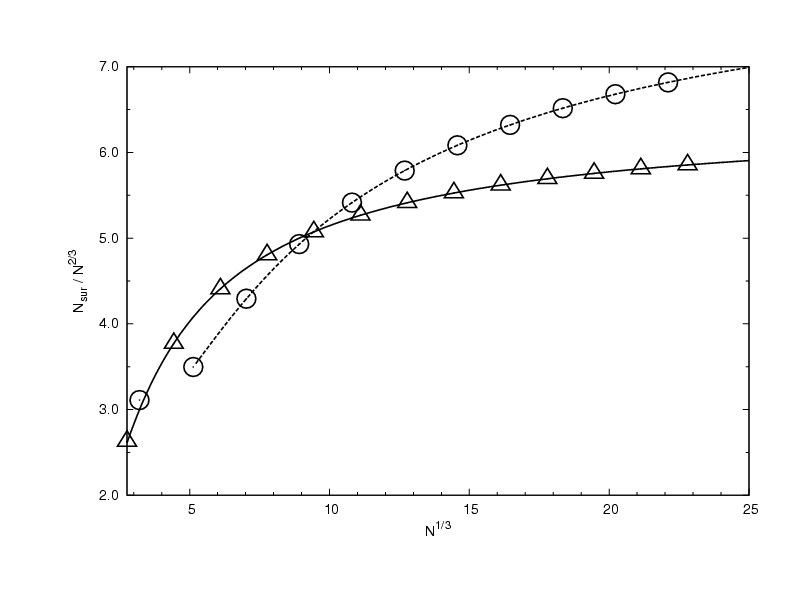}
\caption{\label{fig:4}
Behavior of the ratio $N_{\textrm{surf}}/N^{2/3}$ as function of
$N^{1/3}$, where $N$ is the cluster size, for both the relaxed SC
($\triangle$), and SI ($\bigcirc$) clusters, see
Fig~\ref{fig:3}.\ The actual number of surface atoms of each
stellated cluster, $N_{\textrm{surf}}$, is calculated through the
Lee-Richard algorithm\cite{Lee:1971bx} using a probe radius
$r_{\textrm probe} = 1.66$~\AA.\ The dashed and solid lines are
obtained by fitting the corresponding $N_{\textrm{surf}}$ data to
Eq.~(\ref{eq:6}).\ Note that the smaller cluster of the SI series is
excluded from the fitting.}   
\end{figure*}

\begin{figure*} 
\includegraphics[width=\columnwidth]{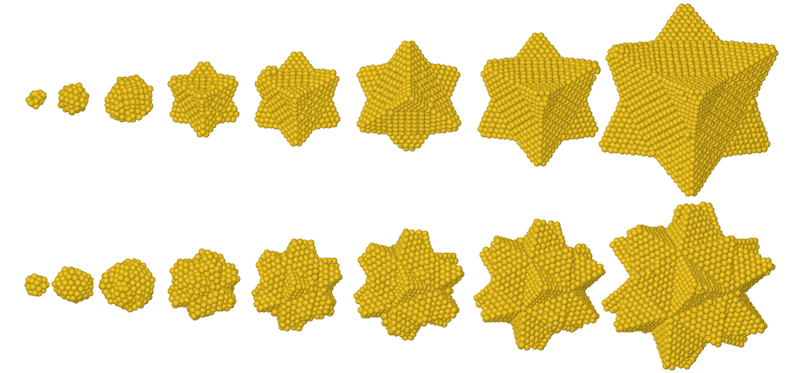} 
\caption{\label{fig:6} 
Same as Fig.~\ref{fig:3}, but for the average equilibrium
configurations of the stellated gold clusters obtained by MDS at room
temperature.\ First row are those gold clusters based on  initial
structures with cuboctahedral core, and second row those with
icosahedral core.\ The first three clusters of each series ($k=1,2,3$)
are structures of low-symmetry or amorphous-like clusters\cite{Garzon1998,Soler:2000uq}.\ The
last cluster in each series corresponds to $N_{\textrm{SI}}(9) = 6
169$ (bottom), and $N_{\textrm{SC}}(10) = 5 631$ (top).}  
\end{figure*}

\begin{figure*} 
\includegraphics[width=\columnwidth]{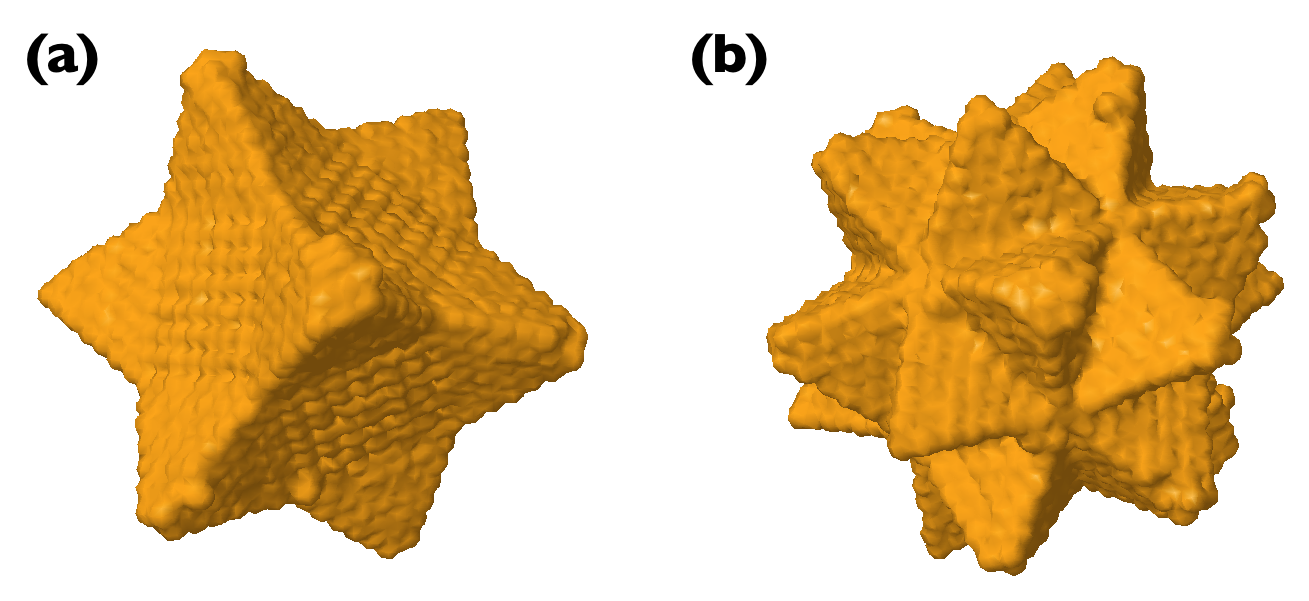} 
\caption{\label{fig:10} 
Snapshots of the equilibrated stellated cuboctahedral and icosahedral  
clusters. In (a) the clusters has 
$N_{\textrm{SC}}(12) = 9 437$ atoms at $T=298~\pm 0.46~{\text K}$.\ The
diameter of the cluster is around $10.8~{\text{nm}}$; in (b) the
cluster has  $N_{\textrm{SI}}(11)~=~10 803$ atoms.\ In this case $T=298~\pm
  0.44~{\text K}$, and the diameter of the cluster is around
  $10.0~{\text{nm}}$. To enhance
surface details a molecular surface representation computed through
the Lee-Richard algorithm\cite{Lee:1971bx} has been used.
} 
\end{figure*}

As it can be observed from the average equilibrium configurations of the two
series of clusters shown in Fig.~\ref{fig:6}, one of the more
evident 
temperature effects on clusters for sizes $347 < N < 9437$ is the
smoothing of all the peaks of the tetrahedral pyramidal motifs of the
relaxed clusters.\  
This is easy to understand due to the natural tendency of lower
coordinated surface atoms to increase its coordination number to the
optimum value, which in this case is driven by thermal energy.\  
For the first three clusters of each series we have found that all the
peaks are completely destroyed in agreement with recent results, which
shown that small gold clusters are low-symmetry or amorphous-like
clusters\cite{Soler:2000uq}.\ By the other hand, for the two largest
clusters of each series, Figs.~\ref{fig:10}, some peaks
remain due mainly to the length of 1 ns used in its equilibration
period.\ Their contribution to the energy could be neglected since, as
we will show below, volume contribution dominates over surface
contributions at these sizes.\  
Eventually, it is expected these peaks are finally smoothed if
equilibration length is extended to 2 ns.\ Thus, the surfaces of our
equilibrated clusters at room temperature for sizes in the range $347
< N \leqslant 10 803$ resembles the TEM images of the gold particles
(a), (b) and (d) in Fig.~\ref{fig:1}.\  

By following the equilibration time history of clusters with sizes
greater than 347 atoms, we have observed in detail how the peaks of
the pyramids are smoothed.  
For any cluster at the initial steps of the equilibration process,
each pyramid tip atom is diffused along one edge and it attaches around
the middle edge for long time, generating metastable states.\  
After a short period, 
these atoms remove (almost always) another atoms from those mid edge positions, 
that are rolled towards  the base of the pyramid, and the total energy is stabilized for longer
periods, which might be considered as an stable state.\  
For the SC clusters, we have found that those atoms form  oscillating
pairs of atoms, which are deposited  
on the surface around vertices connecting square faces.\ 
Nowdays, there are high interest on the collective electronic excitations or
surface plasmons of star-like gold nanoparticles, and these theoretical
findings could give some insights  on the features of the
collective electronic excitations resulting from the interaction of
star-like nanoparticles with the electric field of 
light~\cite{SanchezGil:2010,Feldmann2009,Marzan2008,Hafner2006}.\ 
%
\begin{table}
\caption{\label{tab:4}Same as Table~\ref{tab:2}, but for 
the equilibrated gold clusters with the CR potential model.}
\begin{ruledtabular}
\begin{tabular}{cccc}
    & $\alpha$ & $\beta$ & $\gamma$ \\
\hline

 SC & $6.41  \pm 0.02  $ & $-13.07  \pm 0.33    $ & -- \\
 SI & $8.44  \pm 0.03  $ & $-37.32  \pm 0.50    $ & -- \\

\end{tabular}
\end{ruledtabular} 
\end{table}

\begin{figure*} 
\includegraphics[width=\columnwidth]{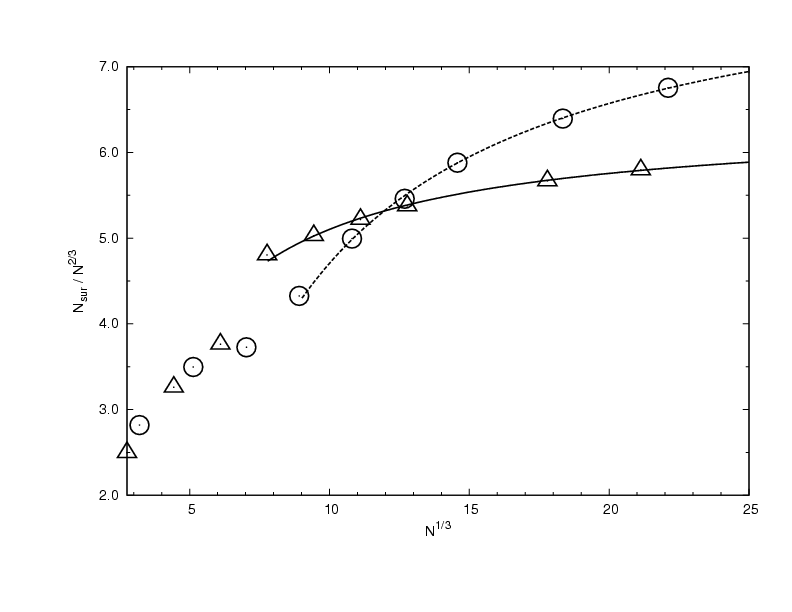} 
\caption{\label{fig:7} 
Same as Fig.~\ref{fig:4}, but for the average equilibrium
configurations shown in Fig.~\ref{fig:6} and Fig.~\ref{fig:10}.\ 
In this case only the $N_{\textrm{surf}}$ data,
for $N\geqslant 469$ has been fitted to Eq.~(\ref{eq:6}) excluding the
coefficient $\gamma$; the first three clusters of each series are
excluded from the fitting, since they are amorphous or low-simmetry
structures~\cite{Garzon1998,Soler:2000uq}.\ 
$N_{\textrm{surf}}$ has also been computed through the
Lee-Richard algorithm\cite{Lee:1971bx} using a probe radius
$r_{\textrm{probe}} = 1.66$~\AA.}  
\end{figure*}

From the above results on equilibrated surface atoms, minimal changes
in the behavior of the number of surface atoms as a function of cluster
size are expected, and this is indeed the case if the curves ${\cal
  N}(N)$ for the SC and SI clusters on Fig.~\ref{fig:4} are compared
with those of the equilibrated clusters shown in Fig.~\ref{fig:7}.\   
The behavior of the corresponding curves, for $N > 347$, is very
similar.\ As in the case of the relaxed clusters described in the
preceding section, curves in Fig.~\ref{fig:7} have been obtained by
linear fitting $N_{\textrm{surf}}$ to Eq.~(\ref{eq:6}), where
$N_{\textrm{surf}}$ is the number of surface atoms of the average
equilibrium configurations shown in Fig.~\ref{fig:6}.\  
The coefficients of the fitting functions for the equilibrated SC and
SI clusters are given in Table~\ref{tab:4}, where the coefficient
$\gamma$ has been excluded from the fitting process in order to avoid
large asymptotic standard errors in the coefficients.\ 
Since the main interest here is on the large stellated structures which
maintain their peaks, the first three equilibrium configurations of
each series have been excluded from the fitting, and also from the
discussions, since they are amorphous or low-symmetry structures
described in detail by Soler {\em et al.}\cite{Soler:2000uq}.\    
>From  Fig.~\ref{fig:7}, the number of surface atoms of the SI clusters is
lower than for relaxed clusters; on the contrary, relaxed and
equilibrated SC clusters have almost the same number of surface atoms.\  
For large clusters, the asymptotic behavior of dotted curves, or solid
curves, is very similar.\ Thus, from the energetic point of view,
details of the surface atoms of very large particles could not be
important since volume atoms dominate energy contributions, but the
smoothed peaks of the clusters can play a significant roll in the
description of the surface plasmons of these nanoparticles.\  

\begin{figure*} 
\includegraphics[width=\columnwidth]{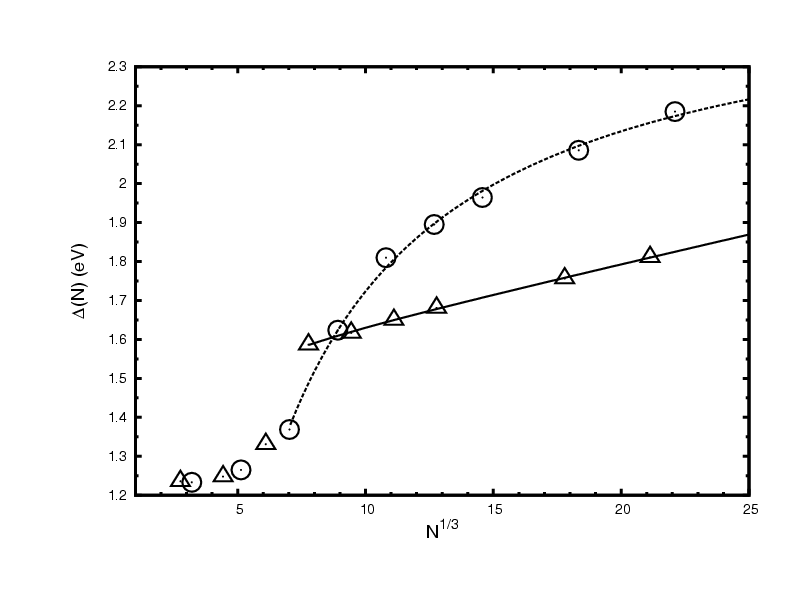} 
\caption{\label{fig:8} Same as Fig.~(\ref{fig:5}), but for the MDS
  results obtained using the CR potential model\cite{Cleri:1993ck};
  the excess energy $\Delta(N)$ has been calculated using the total
  energy.\ Note that we have excluded from the fitting the smallest
  clusters, since they are low-symmetry or amorphous structures with
  non star-like shape.\ 
} 
\end{figure*}

\begin{table}
\caption{\label{tab:5}Same as Table~\ref{tab:3}, but for 
the equilibrated gold clusters with the CR potential model.}
\begin{ruledtabular}
\begin{tabular}{cccccc}
	Cluster type & $A$& $\sigma$\footnotemark[1]& $B$ & $C$ & $D$ \\
\hline


   SC & $-3.6982 \pm 0.0006  $ & $0.0151  $ & $1.496 
  \pm 0.012   $ & --     & $-1.64   \pm 0.54    $ \\
   SI & $-3.7132 \pm 0.0040  $ & $0.0000  $ & $2.545   
\pm 0.107   $ & $-8.22   \pm 0.63    $ & -- \\

\end{tabular}
\end{ruledtabular}
\footnotetext[1]{$\sigma = A - E_{\textrm{ref}}$, 
where $E_{\textrm{ref}} = A_{\textrm{SI}}$.} 
\end{table}
By following the same procedure as for the relaxed clusters, we have
also explored which type of equilibrated gold cluster is more
favorable.\ In this case $E$ in Eq.~(\ref{eq:7}) is the total energy of
a cluster of size $N$, $N_{\textrm{surf}}$ is given by Eq.~(\ref{eq:6})
and Table~\ref{tab:4}, and for $E_{\textrm{ref}}$ we have used
$A_{\textrm{SI}}=-3.7132$\text{ eV/atm} given in Table~\ref{tab:5}.\  
The calculated values for $\Delta_{\textrm{SC}}(N)$ and
$\Delta_{\textrm{SI}}(N)$ are given in Fig.~(\ref{fig:8}) together
with its fitted curves, where the coefficients are given in
Table~\ref{tab:5}.\  
In this case only $E(N)$ data for $N\geqslant 347$ has been fitted to
Eq.~(\ref{eq:9}), and to avoid large asymptotic errors in the
coefficients, $C$ and $D$ have been omitted in the fitting process of
data $E_{\textrm{SC}}(N)$ and $E_{\textrm{SI}}(N)$, respectively.\  
>From interpolation and extrapolation results we have found that the
SC clusters are more favorable in the range $671\lesssim N \lesssim
224~586$, while SI clusters are more favorable out of this range,
excluding the first five smaller clusters.\ Thus, we predict that
stellated gold
clusters coexist around two crossover sizes;  one at $N \simeq
671$ and another one at $N \simeq 224~586$, a result that is 
well agreement with experimental facts (see Fig.~\ref{fig:1}).\ 
 
\section{\label{sec:5}Conclusions} 
We have synthesized and characterized by TEM stellated gold
nanoparticles, as well as performed a systematic study    
for their structure and energetic stability
through geometrical models and by means of molecular dynamics
simulations at room temperature.\ From the analysis of the MDS results,
we can conclude that:  
(a) the morphology and the surface structure of these stellated  clusters for sizes
$N > 347$ are in good agreement with the synthesized NPs, as well as
with the results  
reported recently by Burt {\em et al.}\cite{Burt:2005pd}.\ The structure of
peaks  is maintained and the
coexistence of the two type of particles has been demonstrated.\ 
Smaller clusters are low-symmetry or amorphous structures already studied in
Ref.~\onlinecite{Soler:2000uq}; (b) the stellated gold clusters prefer
SC tructures for cluster sizes in the range $347~<~N~<~226 281$ 
($\approx$ 30 nm), where
surface effects are stronger for SI clusters than for SC clusters; in
fact the behavior of ${\cal N}_{\textrm{SC}}(N^{1/3})$ is almost
lineal.\ We predict that around  a cluster size of 30 nm,  both types of
clusters could coexist, and for larger clusters the SI structure is
more favorable than the SC structure.\ Finally, it is important to emphasize
that we are comparing clusters of different size which have cores of
the same size, cuboctahedral or icosahedral clusters with same magic
size, thus this difference in the number of atoms arises from 
 the tetrahedral motifs attached to their triangular
faces.\

The present study on stellated gold nanoparticles is driven 
mainly by fundamental and applied {\em leit motivs}.\  
From the  fundamental point of view,  
we advanced the hypothesis that stellation on clusters
(Kepler-Poisont solids) is a
process that follows after the nucleation of well defined structures, such
as octahedra and icosahedra (platonic shapes) and cuboctahedra and
truncated octahedra (archimedean solids), where each one of these
shape classifications  could be (althought with not well defined boundaries) 
located in size regions of the  cluster growth. 
From the applied point of view, stellated gold nanoparticles
 have nowadays and projected in the near
future, potential nanotechnological applications in 
disciplines such as the biomedical applications (cell tissue
imaging, sensing and cancer therapy, drug delivery,
etc.)\cite{Prashant2006,Huang2006,Wei2009,Richardson2006}, and
nanophotonics applications\cite{SanchezGil:2010,Feldmann2009,Marzan2008,Hafner2006}.\

\begin{acknowledgments}
This work has been supported by FAI-UASLP (grant C07-FAI-04-21.23), 
PIFI-2008 (grants 24MSU0011E-06 and 24MSU0011E-05), 
and CONACYT grant  106437; the Welch Foundation with grant  number  AX-1615 and  the 
National Science Foundation (NSF)  project number DMR-0830074. We also acknowledge 
Prof. J.A. Alonso (U. Valladolid, Espa\~na) for critical reading and
comments on the manuscript. Finally, authors acknowledge high
performance computational resources granted by the Centro Nacional de
Superc\'omputo (CNS-IPICYT).  
\end{acknowledgments}

\newpage 
\bibliography{AuStellatedClusters_v4}%


\begin{figure*} 
\includegraphics[width=\columnwidth]{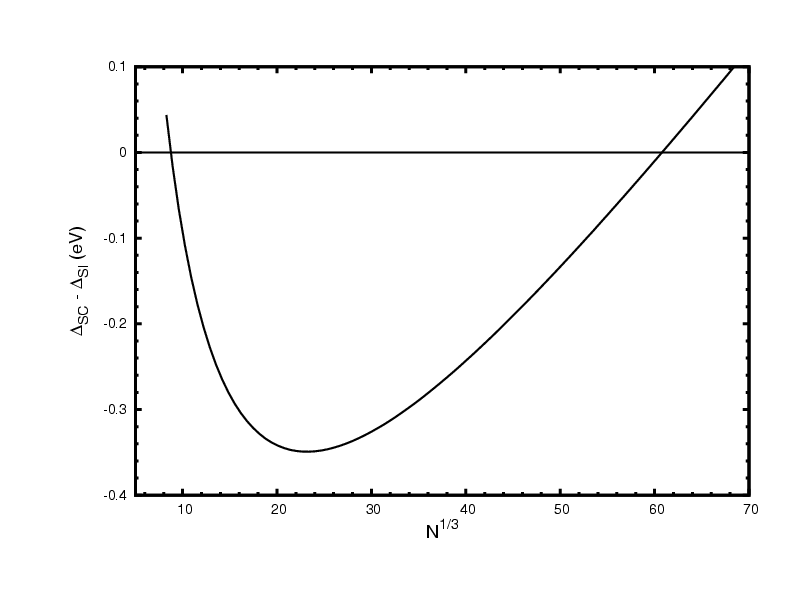} 
\caption{\label{fig:9} 
The difference in excess energy or binding energy per $N^{2/3}$ of the
SC and SI clusters as function of cluster size.\ The curve is obtained
from the fitted curves in Fig.~\ref{fig:7}.\ 
} 
\end{figure*}

\end{document}